\documentclass[aps, 11]{paper}
\usepackage[latin1]{inputenc}
\usepackage{bbm}
\usepackage{graphicx}
\usepackage{cancel}
\usepackage{lmodern}
\usepackage{setspace}
\usepackage{color}
\usepackage{bbm}
\usepackage{amsmath}
\usepackage{amsfonts}
\usepackage{amssymb}
\usepackage{mathrsfs}
\newcommand\beq{\begin{equation}}
\newcommand\eeq{\end{equation}}
\newcommand\bear{\begin{eqnarray}}
\newcommand\eear{\end{eqnarray}}
\usepackage{epstopdf}
\usepackage{pdfpages}
\usepackage{tabularx,ragged2e,booktabs,caption}
\title{Strain shielding from mechanically-activated covalent bond formation during nano-indentation of graphene delays the onset of failure}
\author{\small Sandeep Kumar$^\dagger$ \& David M. Parks $^\ddagger$ \\
\small Department of Mechanical Engineering, Massachusetts Institute of Technology, Cambridge, MA 02139 \\
 \small $^\dagger$lahirisd@mit.edu; $^\ddagger$dmparks@mit.edu \\}
\usepackage[top=2cm,left=2cm,right=2cm,bottom=2cm,foot=1cm]{geometry}
\begin{document}
\maketitle
\begin{abstract}
\small {Mechanical failure of an ideal crystal is dictated either by an elastic instability or a soft-mode instability. Previous interpretations of  nano-indentation experiments on suspended graphene sheets \cite{lee2008measurement, lee2013high}, however, indicate an anomaly: the inferred strain in the graphene sheet directly beneath the diamond indenter at the measured failure load is anomalously large compared to the fracture strains predicted by both soft-mode and acoustic analyses. Through multi-scale modeling combining the results of  continuum, atomistic, and quantum calculations; and analysis of experiments, we identify a strain-shielding effect initiated by mechanochemical interactions at the graphene-indenter interface as the operative mechanism responsible for this anomaly.  Transmission electron micrographs (TEM) and a molecular model of the diamond indenter's tip suggest that the tip surface contains facets comprising crystallographic $\{111\}$ and $\{100\}$ planes.  \textit{Ab initio} and molecular dynamics (MD) simulations confirm that a covalent bond (weld) formation between graphene and the crystallographic $\{111\}$ and $\{100\}$ facets on the indenter's surface can be induced by compressive contact stresses of the order achieved in nano-indentation tests. Finite element analysis (FEA) and MD simulations of nano-indentation reveal that the shear stiction provided by the induced covalent bonding restricts relative slip of the graphene sheet at its contact with the indenter, thus initiating a local strain-shielding effect. As a result, subsequent to stress-induced bonding at the graphene-indenter interface, the spatial variation of continuing incremental strain is substantially redistributed, locally shielding the region directly beneath the indenter by limiting the buildup of strain while imparting deformation to the surrounding regions. The extent of strain shielding is governed by strength of the shear stiction, which depends upon the level of hydrogen saturation at the indenter's surface. We show that, at intermediate levels of hydrogen saturation, the strain-shielding effect can enable the graphene to support experimentally-determined fracture loads and displacements without prematurely reaching locally limiting states of stress and deformation.} \end{abstract}
\hspace*{-0.6cm} Keywords: Graphene, ideal strength, lattice stability, nano-indentation, mechanochemistry, strain-shielding.\\ \\
Lee, \textit{et al.} \cite{lee2008measurement} measured the fracture strength of graphene using nano-indentation experiments and reported an unprecedentedly high intrinsic strength, measuring orders of magnitude greater than those of conventional materials. The nano-indentation involves instrumented indentation of a suspended graphene sheet by a nanoscale diamond indenter up to the point of failure. However, the local stress and strain beneath the indenter are {\em not} directly measurable in such nano-indentation experiments; instead, the indentation load $(F)$ as a function of indentation depth $(u_z\rvert_{r=0})$ during the course of indentation is recorded. Finite Elements Analysis (FEA) based on an appropriate constitutive description of graphene, along with certain kinematic and surface interaction assumptions, is used to simulate the nano-indentation experiment, providing an estimate of the stress and strain beneath the indenter corresponding to the measured fracture load and indentation depth. The inferred local state is taken to represent the intrinsic strength of graphene.\\
%
%
\hspace*{0.5cm}  Using in their FEA model an isotropic nonlinear elastic model containing only up to third-order elastic constants, Lee, \textit{et al.} \cite{lee2008measurement},  estimated the fracture strength of graphene as $\sigma_{f} =130 \pm 10$ GPa and the corresponding areal strain (trace of the 2D logarithmic strain tensor) at fracture as $\epsilon_{a} =0.45$.  Noting that  graphene is elastically anisotropic at large strain,  Wei, \textit{et al.} \cite{Wei2009nonlinear} developed an improved nonlinear constitutive model which incorporates anisotropic elastic response. Employing this model in a FEA simulation \cite{wei2012experimental}, they obtained revised estimates of the fracture strength of graphene as $\sigma_{f} =108 \text{ GPa}$, and of the corresponding fracture strain as $\epsilon_{a}= 0.42 - 0.45$. These simulations indicated that the membrane directly beneath the indenter, where the onset of failure occurs in the simulations, remains in a state of equi-biaxial tension prior to fracture. Yevick \& Marianetti \cite{marianetti2010failure} recently reported that the mechanical strength of a graphene monolayer under equi-biaxial tension is limited by a  soft-mode instability which emerges at  $ \epsilon_a \sim 0.28-0.30$. The breaking strains inferred by Lee, \textit{et al.} and Wei, \textit{et al.} substantially exceed the limiting strain dictated by this soft-mode instability.\\
\hspace*{0.5cm} Wei and Kysar \cite{wei2012experimental} speculated that the soft-mode instability is possibly suppressed in experiments, and that the failure observed in experiments is due to an elastic instability. They assumed that the elastic instability essentially corresponds to the peak of the equi-biaxial Cauchy stress along the loading path. The assumption that  elastic instability corresponds to the peak  Cauchy stress along an assumed deformation path is not always true --- in general, a homogeneously-deformed material can become elastically unstable before reaching this point. In particular, when the unstable eigenmode is orthogonal to the loading path, an elastic instability can precede the maximum in the stress-strain curve \cite{clatterbuck2003phonon, morris2000internal}. By means of acoustic tensor analysis based on a rigorously tested hyperelastic constitutive model, we show that this is indeed the case for graphene. \\
\textbf{Acoustic tensor analysis ---} Both the 2D bulk modulus $\kappa(\epsilon_a)$ and the 2D shear modulus $\mu(\epsilon_a)$ experience a progressive softening with increasing areal strain, ultimately vanishing at $\epsilon_a =0.42$, and $\epsilon_a =0.356$, respectively, as shown in Fig.~[\ref{const}]. Progressive softening of the elastic moduli with dilatant lattice deformation is often responsible for triggering an elastic instability.  Acoustic tensor analysis \cite{Born2, hill1977principles}, which constitutes a useful means of predicting such instabilities, asserts that if, for some pair of unit vectors $\mathbf m$ and $\mathbf n$, 
\begin{equation} \Lambda (\mathbf m, \mathbf n) \equiv (\mathbf m \otimes \mathbf n) : \mathbb A : (\mathbf m \otimes \mathbf n) \le 0,
\end{equation} then the material is elastically unstable, where $\mathbb A = \partial^2 \psi/ \partial \mathbf F^2$ is the acoustic tensor, with $\psi$ denoting the strain energy density per unit area (here, taken to be a function of the logarithmic strain $\mathbf E^{(0)}$), and $\mathbf F$ being the deformation gradient. For 2D equi-biaxial tension,  $\mathbb A$ assumes the simple form:
 \begin{equation} \mathbb A_{ijkl}=
\mathbb L^{(0)}_{ijkl} - \mathbf T^{(0)}_{jk} \delta_{il}, \end{equation} in which $\mathbf T^{(0)} = \partial \psi / \partial \mathbf E^{(0)}$ is the work-conjugate stress tensor,
$\mathbb L^{(0)}_{ijkl} = \partial \mathbf T^{(0)}_{ij} / \partial \mathbf E^{(0)}_{kl}$ is the tensor of work-conjugate tangent moduli, and $\delta_{il}$ is the Kronecker delta.  For this deformation, the plot of $\Lambda_{\mathrm{min}} = \text{min} \, \Lambda (\mathbf m, \mathbf n)$ vs. $\epsilon_a$ indicates an elastic shearing instability ($\mathbf m \cdot \mathbf n =0$) at $\epsilon_a \rvert_{\mu =0} = 0.356$, the point at which the shear modulus vanishes. As shown in Fig.~[\ref{const}], this elastic shearing instability occurs well before the peak in the equi-biaxial Cauchy stress, which is at $\epsilon_a\rvert_{\kappa=0} =0.42$, is reached. \\

\begin{minipage}{\linewidth}
\centering
\includegraphics[scale=0.45]{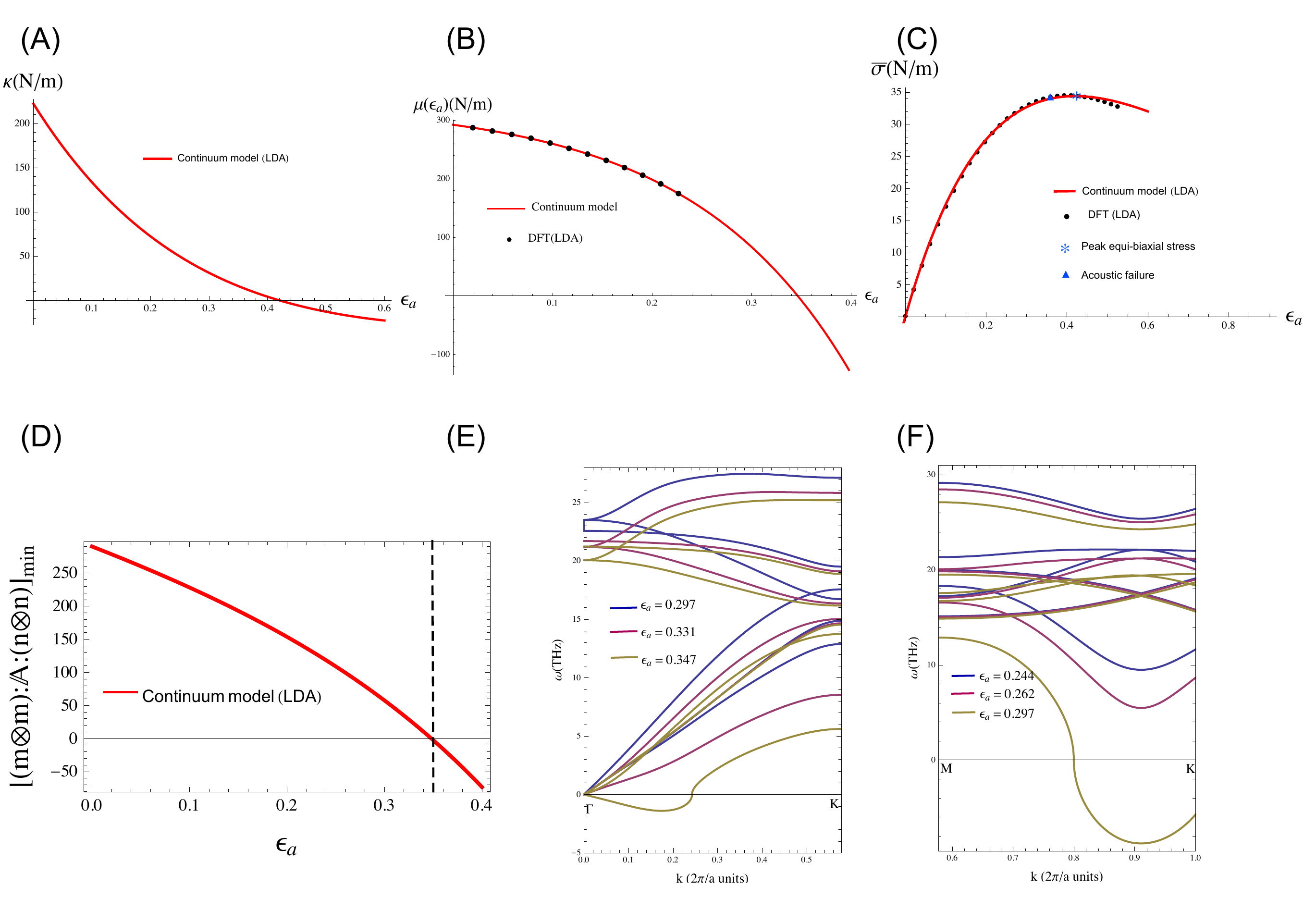}
\captionof{figure}{\small \textbf{Softening of elastic moduli with dilatant area deformation, elastic instability, and soft-mode instability:} Softening of (A)  the 2D bulk modulus and (B) the 2D shear modulus with increasing areal strain $\epsilon_a$. (C) Plot of  equi-biaxial Cauchy stress $\bar{\sigma} = \text{tr}\,\pmb\sigma/2$ vs. $\epsilon_a$. (D) For equi-biaxial strain, acoustic tensor analysis predicts an elastic instability at $\epsilon_a =0.356$. (E) Independent phonon calculations confirm a shearing elastic instability at $\epsilon_a =0.35$ through a vanishing slope of the long-wavelength TA dispersion relation. (F) Well before this elastic instability, a soft-mode instability occurs at $\epsilon_a =0.297$ \cite{marianetti2010failure}. }
\label{const}
\end{minipage} \\ \\
\textbf{ Phonons-based failure function ---} Another  mechanism by which a crystalline material can fail mechanically is a soft-mode instability \cite{marianetti2010failure, clatterbuck2003phonon}. While acoustic tensor analysis captures only elastic instabilities, a phonons-based stability analysis detects both elastic and soft-mode lattice instabilities. Employing a phonons-based stability analysis, we construct a failure criterion for graphene which provides an analytical description of lattice instabilities of all kinds over the set of all deformations. The scalar-valued failure function $\mathcal F (\epsilon_{\text{max}}, \epsilon_{\text{min}}; \theta)$--- is defined in terms of two principal logarithmic strains $\epsilon_{\text{max}}$ and $\epsilon_{\text{min}}$ and the principal stretch direction $\theta$. When graphene is subjected to a homogeneous  strain for which $\mathcal F>0$,  the underlying lattice is stable w.r.t. perturbations; conversely, when $\mathcal F<0$, the lattice is unstable, and the transitional condition $\mathcal F = 0$ parametrizes material deformed to incipient lattice instability (see \cite{Kumar2} for details). \\ \\ 
\textbf{FEA simulation of nano-indentation ---} Employing the constitutive model and the failure criterion in a FEA simulation of nano-indenation of a graphene sheet supported over a substrate with a micro-cavity, we analyze  failure of the graphene sheet while adopting the same geometric parameters, kinematic boundary conditions, and contact models used previously \cite{wei2012experimental}  to model the experiments in \cite{lee2008measurement, lee2013high}.
%
%
%
 By monitoring strain as a function of indentation depth $u_z \rvert_{r=0}$  in the simulation with tip radius $\rho=16.5\text{ nm}$, two points of potential failure are predicted and shown in Fig.~[\ref{res}-C]. The first ---  occurring at  $u_z \rvert_{r=0} =90 -92 \text{ nm}$ ($\epsilon_a \rvert_{\text{soft-mode}} \sim 0.28-0.30$)--- is a soft-mode instability; while the second ---  occurring at $u_z \rvert_{r=0} =100-102 \text{ nm}$ ($\epsilon_a  \rvert_{\text{acoustic}} =0.356$ ) ---is an elastic instability (the one arising from the vanishing shear modulus).  In contrast, the experimental failure did not occur until 
$u_z \rvert_{r=0} = 109 \text{ nm}$. \\ \\
\begin{minipage}{\linewidth}
\centering
\includegraphics[scale=0.6]{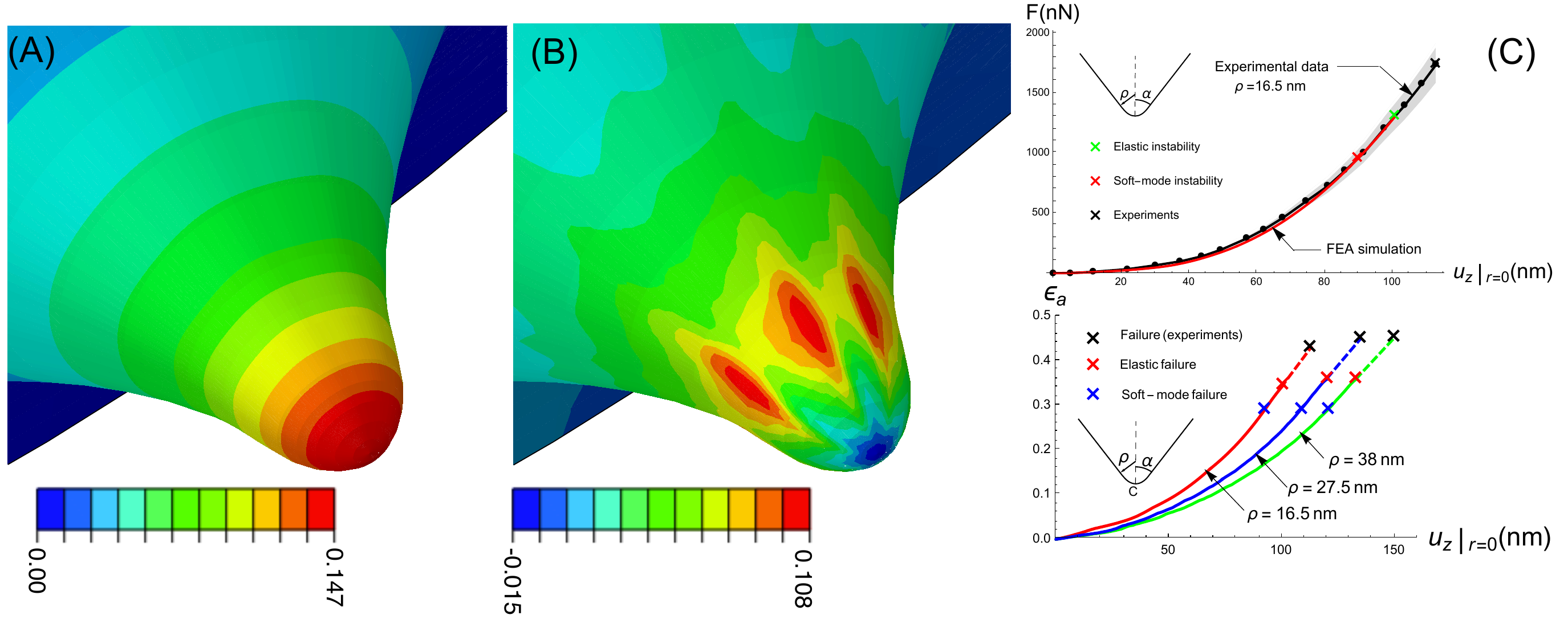}
\captionof{figure}{\small \textbf{Simplified finite element analysis of nano-indentation:} Contours of  (A) maximum principal strain $\epsilon_{\text{max}}$ and (B) failure function $\mathcal F$ beneath the indenter at the indentation loading corresponding to incipient soft-mode failure (blue color at base indicates spatial focus of the instability). (C) \textit{Top}---  load vs. indentation response  from experiment and simulation for indenter tip radius $\rho =16.5 \text{ nm}$ and flake radius $a=0.5\, \mu\text{m}$. Also shown are inferred failure states corresponding to the elastic and the soft-mode instabilities, and the  failure state measured by Lee, \textit{et al.}. \textit{Bottom}--- Areal strain beneath the indenter vs.  indentation depth $u_z\rvert_{r=0}$ for three different indenter root radii used in experiments \cite{lee2008measurement, lee2013high}. Values of $\epsilon_a$ corresponding to elastic and soft-mode failures are marked by red and blue crosses ($\times$), respectively. Dashed lines are extrapolated  to respective experimental depths at failure.}
\label{res}
\end{minipage} \\
Thus we remain confronted by the previously-noted anomaly: while  FEA simulations based on an appropriate constitutive model for graphene, along with certain kinematic boundary conditions, geometric variables, and surface interaction models, predict experimental load vs. indentation extremely well, the experimentally-observed failure occurs much later than predicted by either of the two lattice instabilities.  Moreover, these conclusions also apply to simulations of additional experiments conducted using nanoindenters of different root radii; while all the $F- u_z\rvert_{r=0}$ relations are well-predicted, curves of
$\epsilon_a$ vs. $u_z\rvert_{r=0}$, shown at the bottom of Fig.~[\ref{res}-C], all predict the occurrence of respective soft-mode and elastic instabilities well before experimentally-observed failures.  Resolution of these discrepancies is important because it is the key to obtaining a more precise assessment of the ideal strength of graphene, as well as a correct identification of the mechanisms governing the failure.
 
Ambient-temperature MD simulations of nano-indentation of a suspended graphene sheet by a hydrogen-passivated diamond tip (discussed in detail below) show that the fracture beneath the indenter indeed initiates at $\epsilon_a  \rvert_{\text{MD}}\sim 0.28- 0.30$,  the same value at which the soft-mode instability predicted by the phonons-based failure criterion is noted in the FEA simulations. The hydrogen-passivated surface  ensures that the tip interacts weakly with the graphene sheet throughout the course of indentation. This simulation result supports the proposition that a failure prediction based on the soft-mode instability in graphene remains a relevant indicator of failure in nano-indention, even in the presence of complicating effects including the proximity of the indenter's surface and thermal fluctuations at finite temperatures. The atomic motions taking place during the zone boundary soft-mode are locally in-plane \cite{marianetti2010failure}, to which the non-bonded vdW interaction does not impose significant resistance; 
 thus, concurrence of the phonon-based failure criterion with the results of this MD simulation could be expected, since below the Debye temperature, phonons constitute a complete normal basis of lattice vibrations. 
   
 In an effort to identify and understand the origin of the discrepancy in failure prediction, we investigate the roles of  simplifications and assumptions, other than the graphene constitutive response, embedded within the FEA simulations, as these can potentially affect the details  of the simulation results and hence the interpretation of experiments.  Three critical modeling assumptions can directly affect the inferred failure conditions observed in nano-indentation simulations: (1) the assumed kinematic boundary condition at the graphene boundary; (2) simplifications related to the indenter shape and size; and (3) the simplifying assumptions regarding  interaction at the graphene-diamond interface. \\
\textbf{Boundary conditions ---} For sufficiently small indentation loads,  adhesive and frictional forces from the substrate act to `clamp' the graphene sheet  at the cavity-periphery, $r=a$. For  higher indentation loads, the supported graphene external to the cavity can slide radially inwards while stretching, drawing  a surrounding annular region into the cavity; in such cases, the radial position of  the effective `clamped' boundary shifts to $r>a$ \cite{kitt2013how}.  Intuitively, clamping at $r>a$ should distribute strain over a larger extent of the material, tending to reduce strain beneath the indenter. However, this also results in excessive softening of the entire load-indentation curve (see Fig.~[S12B] in SM), leaving it in substantial disagreement with experimental data. Thus, we concur that for the range of loads in the experiments, $u_r\lvert_{r=a}=0$ is an appropriate boundary condition.\\
\textbf{Indenter size and shape ---} Results shown in Fig.~[\ref{res}-C] (bottom) confirm the intuitive notion that larger tip radii spread contact stress
over larger areas, reducing strain beneath the indenter and delaying inferred failure to greater indentation depths.
 However, for the sharpest tip radius, $\rho$ = 16.5 nm, even with a  20\% larger tip radius, the indentation depth at which a soft-mode instability is first encountered  increases by no more than $2 \text{ nm}$ (see Fig.~[S12A] \& Fig.~[S11A]). To minimize inaccuracies arising from a `spherical cap' idealization of the indenter-geometry, we performed FEA simulations in which the indenter geometry is directly sampled from a high-resolution TEM image of the tip (see Fig.~[S11B] in SM), and again found negligible differences in the loading conditions at which a local soft-mode was first encountered. We conclude that neither a nominal error in tip radius measurement nor assumptions related to the indenter-shape can be possibly resolve the anomaly in inferred failure strain.\\
\textbf{Graphene-tip vdW interaction ---} Under most circumstances, the interaction between nano-scale contact systems is predominantly the van der Waals (vdW) interaction, arising from instantaneous charge fluctuations  \cite{ sandoz2013role, sandoz2012atomistic, lee2010frictional}. We have heretofore ignored  adhesive interfacial forces in the simulations, and, following Lee, {\textit et al.} and Wei and Kysar, have assumed a frictionless, non-adhesive hard contact between graphene and the indenter.
To determine if adhesive vdW forces play a role in delaying inferred failure, we performed FEA simulations incorporating  vdW-adhesion at the graphene-tip interface. However, the inferred failure load and depth remain the same as in the simulations without  interfacial vdW adhesion. We also note that  the relatively small friction attributed to graphene-tip contact (assumed friction coefficient $\mu_f =0.01- 0.02$, see Sandoz, \textit{et al.} \cite{sandoz2012atomistic}) do not substantially affect results, either. \\

\noindent
\textbf{Mechanically-activated covalent interaction ---} Notably, in both FEA  and MD simulations that assume weak vdW interaction at the interface,  no residue accumulates on the indenter tip in the post-failure regime.  Had the only interaction  between the two surfaces  indeed been vdW,  this might be expected, since  vdW forces  have a negligibly small tangential-stiction component (see SM).  However,  post-indentation TEM  images from nano-indentation experiments \cite{lee2008measurement, rasool2013measurement} {\em do} show graphene residue on the tips. For example,  Lee \textit{et al.}'s post-indentation image of the indenter,
shows an apparently well-adhered accumulation of graphene-residue on a localized portion of the tip (Fig.~[\ref{stickind}-B]). The axis of the tip is aligned along a $\langle 100 \rangle$-crystallographic direction of diamond, and the end-facet of the tip is likely  a \{100\}-plane  \cite{koyama2012probe}. The normal vector to the tip surface region exhibiting graphene accumulation makes an angle near $55^\circ$ with the indenter axis. One such direction is a $\langle111\rangle$- crystallographic axis of diamond, suggesting that the adhering facet may be a \{111\}- crystallographic plane.  A similar accumulation pattern is observed in Rasool \textit{et al.}'s post-indentation TEM images as well (see Fig.~[\ref{stickind}-A]). The localized and well-adhered pattern of the graphene-residue accumulation on the indenter suggests that an interfacial interaction much stronger than vdW forces, such as of covalent nature, may  possibly be active at the interface. Previously, such mechanically-activated covalent bond formations have been observed in atomic force microscopy (AFM) experiments \cite{erlandsson2000force, beyer2005mech,jarvis1996direct}. \\ \\ 

\begin{minipage}{\linewidth}
\centering
\includegraphics[scale=0.45]{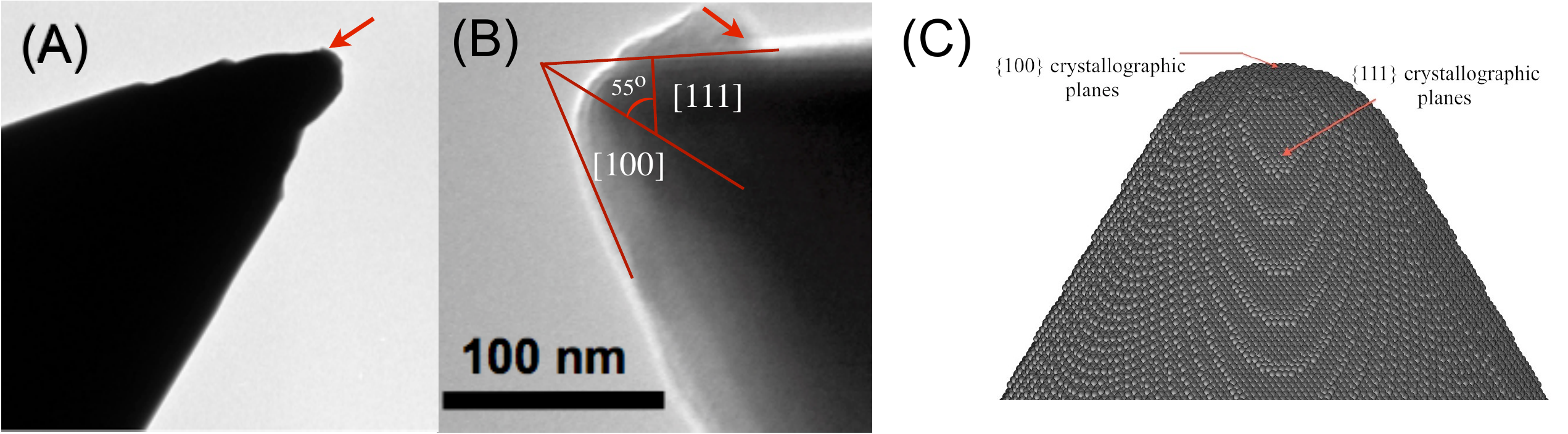}
\captionof{figure}{\small \textbf{Post-indentation TEM image of diamond indenter-tip}: Post-indentation images of the indenter (A) from Rasool, \textit{et al.} \cite{rasool2013measurement} and (B) from Lee, \textit{et al.} \cite{lee2008measurement}. In both images,  graphene residue accumulated on the tip, shown by a red arrow, should be noted.  (C) Molecular model of a spheroconical nanoindenter with $\rho=16.5 \, \text{nm}$, made of single-crystalline diamond. 
Note the \{100\} and \{111\} crystallographic planes exposed on the surface.}
\label{stickind}
\end{minipage} \\ \\
%
 We now address the possibility of mechanochemical interactions at the graphene-diamond \{111\} and the graphene-diamond \{100\} interfaces during nano-indentation. To this end,  we perform dispersion-corrected DFT calculations simulating a graphene monolayer  forced against a \{111\} diamond surface. Calculations are carried out for several different registries, i.e., laterally-shifted configurations of the graphene lattice w.r.t. the diamond lattice such that an atom in the graphene lattice, originally at $m \mathbf a_x + n \mathbf a_y$,  is shifted to $m \mathbf a_x + n \mathbf a_y + \pmb \eta$; $\pmb\eta = (\eta_x, \eta_y)$ is  the shift vector.  All the registries considered exhibit two distinct local minima (wells) in the adsorption energy ($\psi_{\mathrm{adh}}$) vs. separation ($\zeta$) landscape (see Fig.~[\ref{vdWfc}]). 
  %
 %
 %
 %
 The common broader well, centered at $\zeta \doteq 2.78 \mathring{A}$, corresponds to the physisorbed state of graphene on the diamond surface --- in which  interaction is mainly due to interfacial vdW forces, and no electronic charge transfer is involved.  The narrow wells, at smaller separations, correspond to  chemisorbed states --- marked by formation of chemical bonds between graphene and the diamond surface. Bonding is confirmed by the charge density contours at the graphene-diamond interface, as mapped at various separations. Substantial charge overlap between the graphene monolayer and the (111) diamond surface, which is absent at large separations $\zeta > 2.0 \mathring{A}$, occurs at  $\zeta \approx 1.45 -1.75 \mathring{A}$. Fig.~[\ref{vdWfc}-B] shows that the activation stress for reaching the bonded well is within the range 1- 3 GPa, while Fig.~[\ref{vdWfc}-C] (bottom) confirms that  contact stresses of this magnitude are generated over growing portions of the $16.5 \, \mathrm{nm}$ tip as indentation  increases. 
 MD simulations, performed with a reactive AIREBO potential \cite{stuart2000reactive}, also confirm the occurrence of stress-assisted bond formation at graphene-diamond \{111\} and graphene-diamond\{100\} interfaces (see Figs.[S15-S21] in SM). 
 \\
 
\begin{minipage}{\linewidth}
\begin{center}
 \includegraphics[scale=0.6]{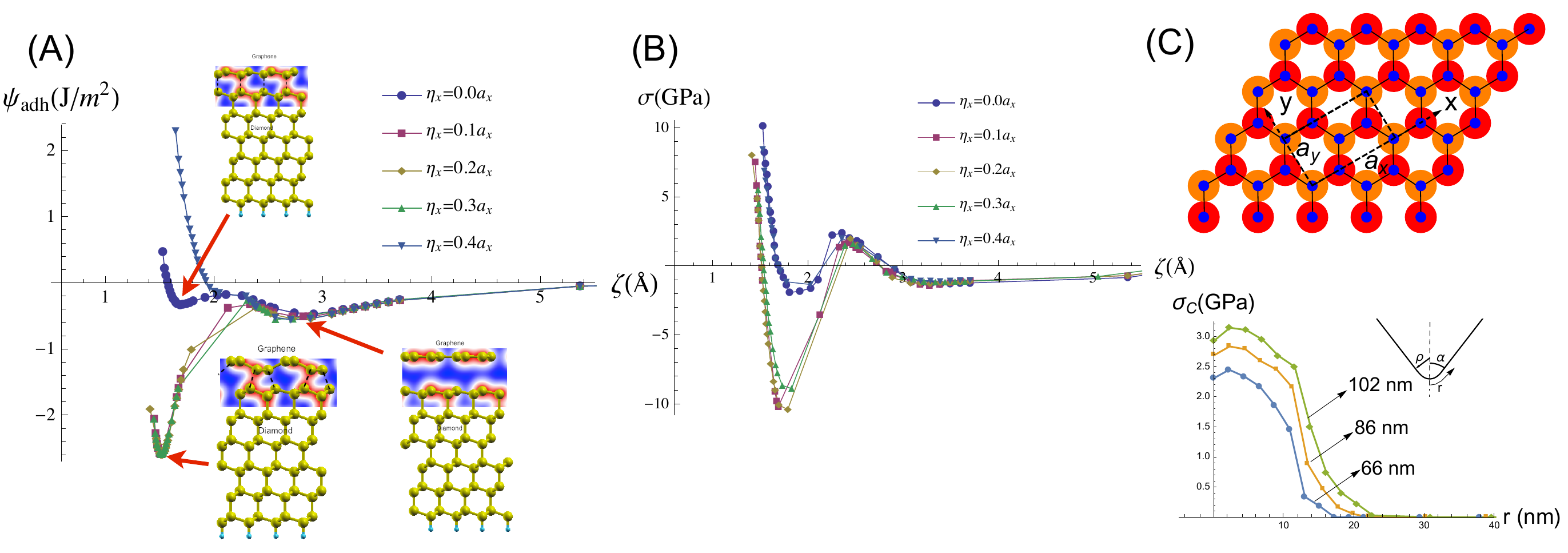} \\
\end{center}
\captionof{figure}{\small \textbf{Graphene-diamond \{111\} interaction as a function of separation:} (A) Plots of $\psi_{\mathrm{adh}}$ vs. $\zeta$ for different registries (see also Figs. [S2 \& S3]).  Also shown are relaxed configurations and  charge density contours for certain chemisorbed and physisorbed states of graphene on diamond. 
 (B)  Contact stress ($\sigma$) vs. $\zeta$ for different registries.   (C) Top --- Schematic of graphene on top of diamond \{111\} surface (blue atoms: graphene lattice; red and orange atoms:  top and second layers of diamond, respectively). \textit{Bottom} --- FEA contact stress calculated along a radial line at various indentation depths. }
\label{vdWfc}
\end{minipage}
\\ \\
The shear strength of a planar C-C interface  depends strongly on the fractional hydrogen-coverage, $\theta_H$, at the interface \cite{zilibotti2011ab}.  Two limiting cases are: (1) fully-hydrogenated state,  $\theta_H =1$, in which case purely vdW interaction holds at all values of contact stress; and (2) fully-clean state,  $\theta_H=0$, in which case the interaction is vdW at low contact stresses, but becomes strongly covalent once the interfacial contact stress  reaches the activation stress. A major distinguishing factor between the two kinds of interaction is the resulting shear stiction, $\tau_0$: a purely vdW interaction possesses a negligible $\tau_0$-value, while  covalent interaction is associated with a large shear strength. For intermediate values of $\theta_{H}$, $\tau_0$  decreases with  increasing $\theta_H$ \cite{zilibotti2011ab}. The surface hydrogen coverage on a diamond surface is determined by its process history  \cite{pate1986diamond, evans1977chemical}. For example, experiments suggest that an as-cleaved diamond surface is nearly clean, and upon annealing, it becomes hydrogenated \cite{kawarada1996hydrogen}; while an ablative treatment such as ion beam milling --- which is often involved in fabrication of nano-indenters \cite{koyama2012probe} --- removes some  hydrogen from the diamond surface, leaving it  in a partially-hydrogenated state \cite{evans1977chemical}. \\
\textbf{Mechanochemical FEA simulation ---} We now examine the consequences of bonding-induced shear stiction at the graphene-indenter interface by means of FEA simulations which duly incorporate mechanochemical details of the graphene-indenter interaction. These simulations  allow the development of mechanically-activated covalent interface interactions with graphene at  indenter tip locations corresponding to C\{111\} and  C\{100\} facets. 
 Based on the picture emerging from these simulations, we identify a strain-shielding mechanism which controls the dispersion of strain beneath the indenter, as illustrated in Fig.~[\ref{mech}]. Initially,  interaction between the two surfaces is dominated purely by vdW forces, since the contact stress beneath the indenter at small indentation is insufficient to initiate bonding. During this phase,  strain build-up mostly occurs directly beneath the indenter (Fig.~[\ref{mech}-A]). As  indentation continues,  bonding between graphene and diamond initiates at the base of the indenter, once the requisite contact stress is reached. The shear-stiction associated with this covalent interaction inhibits the subsequent stretching of the graphene sheet in this region, and the unbonded surrounding region undergoes a little extra strain instead. Effectively, the cumulative action of the shear stiction can be perceived as a mechanism that transfers  strain from the bonded region to the surrounding region,  creating  a 
 strain-shielded region at the base of the indenter (Fig.~[\ref{mech}-B]).  At even larger depths/loads, as graphene wraps around the indenter,  covalent interaction with lateral \{111\}-facets of the diamond indenter is also initiated, and  the associated shear-stiction again transfers the strain to nearby regions (Fig.~[\ref{mech}-C]). The process is illustrated in
 Fig.~[\ref{mech}-D], which contrasts the evolution of maximum principal strain at two near-tip locations for vdW and induced covalent interaction. \\ \\
\begin{minipage}{\linewidth}
\centering
 \includegraphics[scale=0.5]{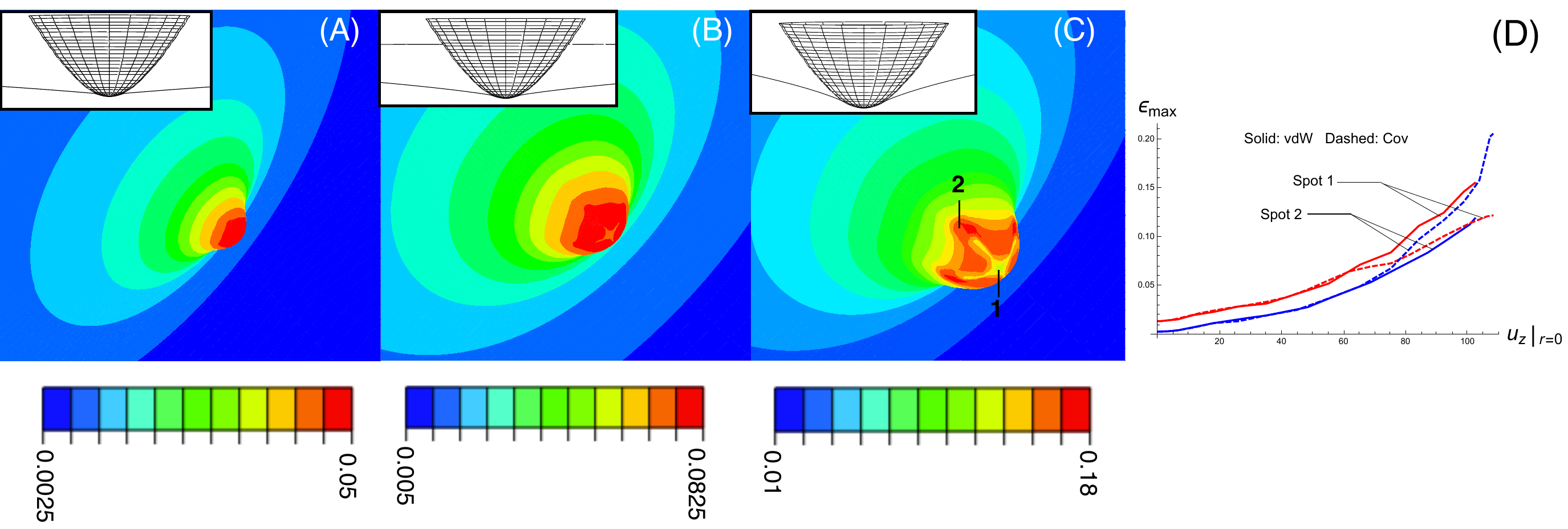}
\captionof{figure}{\small \textbf{Strain-shielding mechanism :} Contours of maximum principal strain at various stages of indentation. (A) $u_z\rvert_{r=0}= 55 \text{ nm}$:  interaction between graphene and diamond tip is purely vdW, and  maximum strain occurs at the point beneath the indenter. (B) $u_z\rvert_{r=0}= 75 \text{ nm}$: formation of a strain-shielded zone indicates that a mechanochemical interaction has been initiated  at the graphene-indenter interface in the region beneath the indenter. (C) $u_z\rvert_{r=0}= 107.5 \text{ nm}$:  As  contact area grows with indentation,  \{111\}-facets on the indenter tip also  interact covalently with the graphene sheet, as indicated by formation of strain-shielded zones in those regions. (D) Maximum principal strain at material points `1' (in red) and `2' (in blue) vs. $u_z\rvert_{r=0}$ for induced covalent interaction (dashed lines, this simulation) and vdW interaction (solid lines, Fig.~[\ref{res}]).}
\label{mech}
\end{minipage} \\ \\ 
By varying the magnitude of $\tau_0$ in FEA simulations, we  explore  conditions for optimal strain shielding that    redistributes strain and delays the onset of instability to a larger load/depth (see Tab.[S5] in SM). 
For an intermediate range of $\tau_0$-values corresponding to a partially-hydrogenated state of the diamond surface, $\theta_H \sim 70\%-80\%$, optimal strain transfer occurs, spreading the strain to the maximal area and limiting the strain-intensification beneath the indenter (Fig.~[\ref{results}-A]). \\

This delays the onset of instability, allowing the inferred failure to occur at an indentation depth/load close to the 
experimentally-measured values (Fig.~[\ref{results}-C]). The  load and indentation at inferred failure in the simulations are  $F = 1540 \text{ nN}$, and $u_z\lvert_{r=0} = 107.5 \text{ nm}$, respectively --- which are within the $\pm 9.5\%$ and  $\pm 3.5\%$  ranges reported for the measured values, $1742$ nN and $109$ nm, respectively. Based on the contours of the failure function $\mathcal F$,  failure is inferred to initiate away from the center, just beyond a bonded region (Fig.~[\ref{results}-B]); such a failure initiation site is consistent with the location of retained residue shown in
Fig.~[\ref{stickind}]. \\ \\
%
%
\begin{minipage}{\linewidth}
 \centering
\includegraphics[scale=0.55]{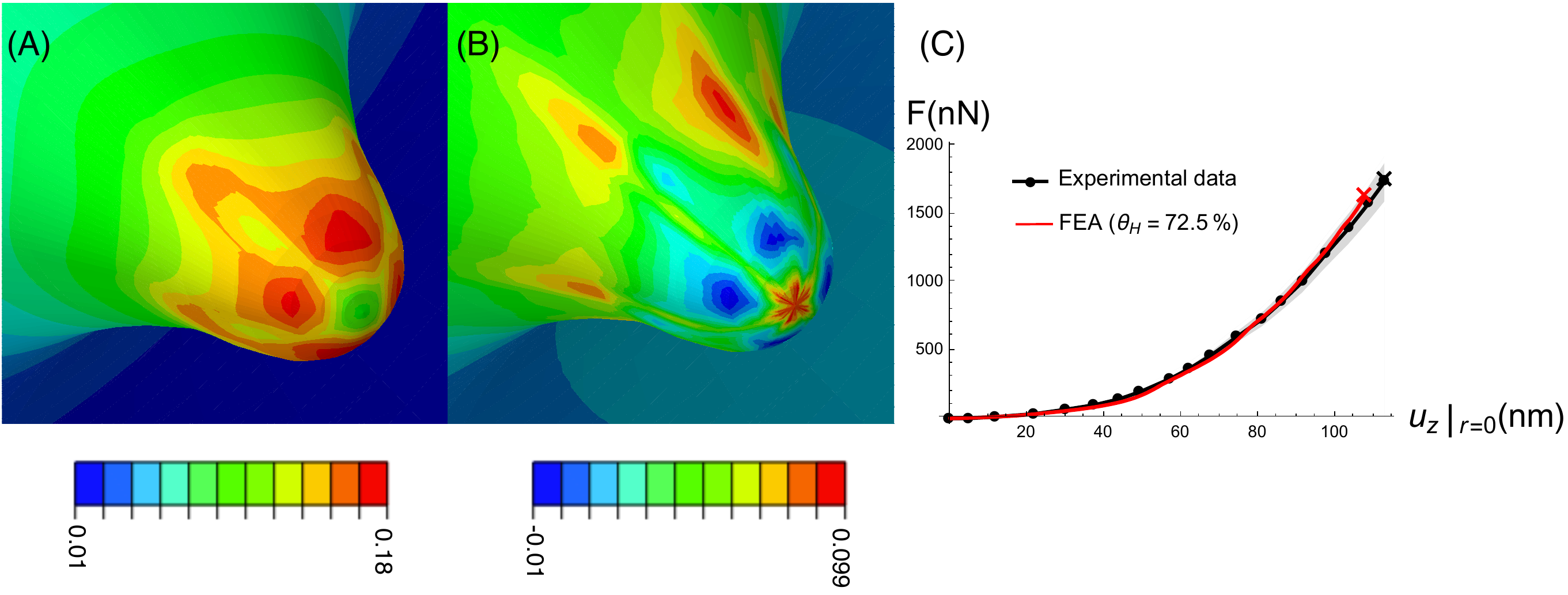}
\captionof{figure}{\small \textbf{FEA simulation allowing development of a covalent interaction at graphene-indenter interface:} Contours of (A) the maximum principal logarithmic strain and (B) the failure function $\mathcal F$ beneath the indenter prior to failure. Note the redistribution of strain in (A) due to graphene-diamond interaction, limiting strain intensification beneath the indenter. Blue contours in (B) show regions where $\mathcal F <0$, i.e., the material has reached incipient failure. (C) Corresponding load vs. indentation depth.}
\label{results}
\end{minipage} \\ \\
In particular, we note that the strength of shear stiction is the key variable governing the performance of this mechanism. Neither a too-small nor a too-large stiction is conducive for maximal dispersion of strain. Too small a stiction (as in the case of fully-hydrogenated diamond) induces no strain transfer, and strain rapidly concentrates in a small region directly beneath the indenter.  Conversely, too large a stiction (as in case of bare diamond), heavily shields the bonded region beneath the indenter, 
%
%
 resulting in a rapid strain buildup within annular sectors in the vicinity of the bonded region (see Tab.[S4] in SM). \\ 
 \textbf{MD simulations ---} MD simulations of nano-indentation are carried out using AIREBO potentials of Stuart \textit{et al.} \cite{stuart2000reactive}, which 
  capture major features of the mechanochemical C-C interactions. Three levels of hydrogen saturation on the indenter surface are considered: bare  ($\theta_H =0\%$); fully-hydrogenated  ($\theta_H =100\%$); and 
 partially-hydrogenated ($\theta_H \approx 60\%$). For all  cases considered, a one-to-one qualitative agreement between FEA and MD is noted --- (1) while both large and small hydrogen coverages result in early failures, an intermediate level coverages requires the largest indentation loads to initiate failure in the graphene. (2)  As in FEA simulations (Fig.~[\ref{mech}-D]), MD simulations also indicate  strain-shielding beneath the indenter, as shown in Fig.~[\ref{MDresults}-D]. (3) Post-fracture atomic trajectories from MD confirm when strong covalent interaction occurs, e.g., $\theta_H =0$ and $\theta_H=0.6$, a residue accumulation is noted, while it is absent when $\theta_H=1$, for which the graphene-tip interaction is weak, as previously noted. \\
 %
 %
 \begin{minipage}{\linewidth}
 \centering
\includegraphics[scale=0.5]{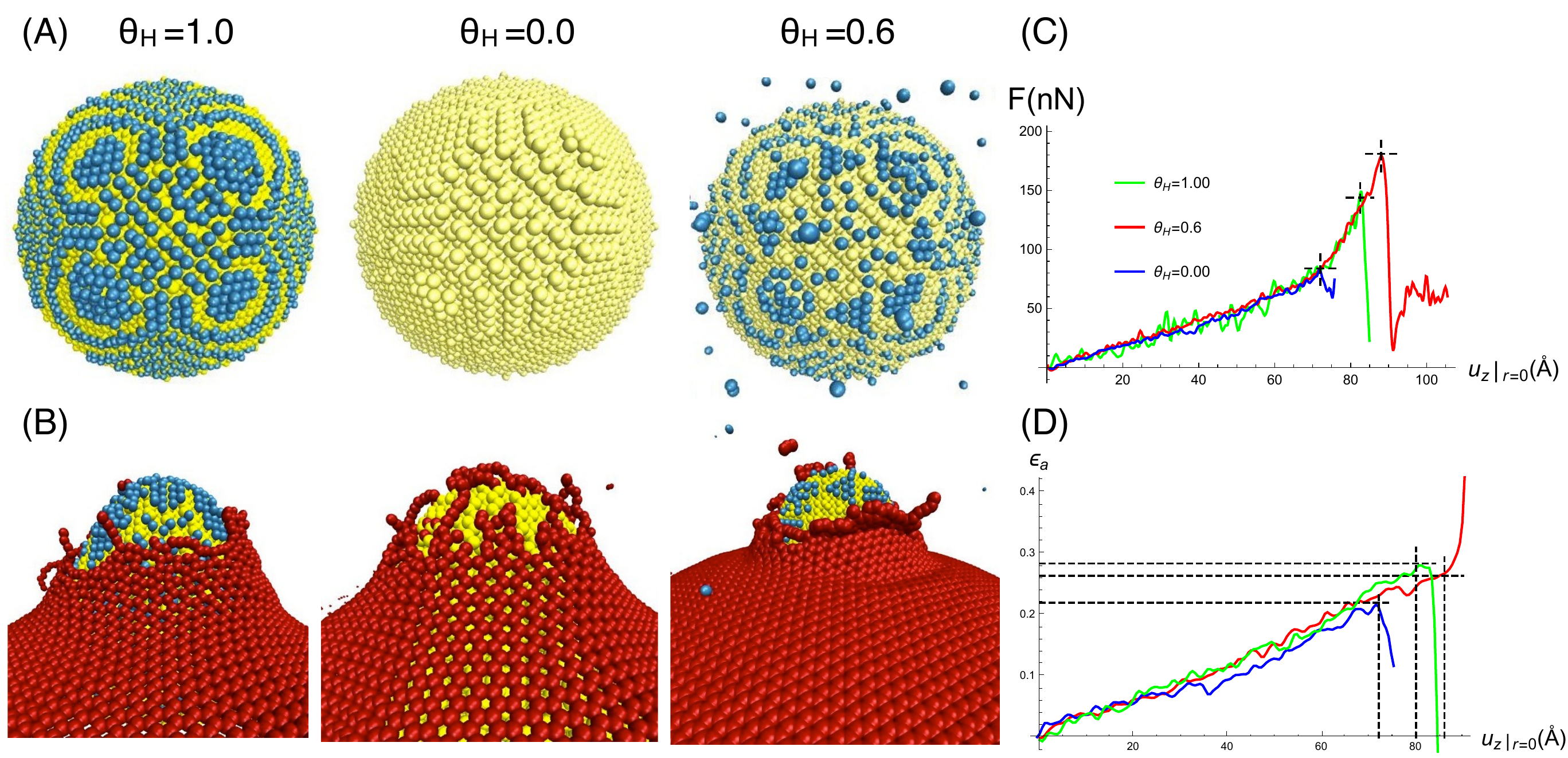}
\captionof{figure}{\small \textbf{MD simulations of graphene indentation for three levels of hydrogen saturation of the diamond indenter:} (A) Illustration of varying hydrogen saturations of the indenter surface. Carbon atoms of the diamond tip in yellow, with hydrogen in blue.
(B) Post-fracture images for the three cases, with graphene carbon atoms in red. (C) Force vs. indentation curves, with the points of failure marked by $+$, and (D) areal strain beneath the indenter vs. indentation depth for the three cases.}
\label{MDresults}
\end{minipage} \\ \\ 
\textbf{\small Conclusion ---}
We have shown that the high local contact stresses generated in nano-indentation can induce strong covalent interaction at crystallographically-specific portions of the graphene-indenter interface, with consequences that can dramatically affect the measured failure load in such experiments. It is a significant finding, since contact interactions in  nano-indentation experiments on graphene have previously been assumed to be weak, and their roles in defining failure modes were implicitly ignored.
As a result, in previous interpretations of  nano-indentation experiments, the inferred failure strains in graphene have appeared to be inexplicably large from the perspective of lattice stability analysis. Performing numerical simulations at multiple scales, we shed light on the mechanistic processes activated by the mechanically-induced interfacial covalent interaction, principally an alteration of the strain landscape in graphene beneath the indenter, and explain the apparent delay in inferred failure in a fashion consistent with lattice stability analysis.\\

 The nano-indentation of two-D materials, such as graphene, constitutes a special case. In the nano-indentation of defect-free bulk-crystals, the ensuing lattice instability  --- the mechanism governing incipient plasticity --- emerges in the interior, removed from the indenter's direct influence \cite{li2002atomistic, tadmor1999nano}. In contrast, for structurally 2D materials, the potentially-unstable regions remain in intimate contact with the indenter surface throughout the course of indentation, and, as a result, the mechanochemistry of the membrane-indenter interface can play a significant role in determining local critical conditions.
For this reason,  measured failure loads/indentations for such 2D materials cannot always be straightforwardly
translated into limiting material parameterizations (strength; critical strain) without attending to potential mechanochemical interactions. The methodologies presented here constitute a multi-scale framework for assessing the mechanochemistry of the membrane-indenter interface and its potential to influence critical local conditions.  Of course,  mechanochemical details of indenter-membrane interactions vary from material to material; however, the framework presented here is general, and should be broadly applicable to nano-scale contact experiments on other monolayer materials, e.g., h-BN, silicene, and two-D analogues of bulk crystals, such as $\text{MoS}_2$, as well.

\end{document}